\documentclass[aps,preprint,amssymb,12pt,floatfix]{revtex4}
\usepackage{subfigure}
\usepackage{graphicx}
\usepackage{rotating} 
\usepackage{color}
\usepackage{rotating}
 \usepackage{amsmath,amsthm}
\usepackage{epstopdf}
\usepackage{soul}
\begin{document}
\title{Iterative Annealing Mechanism for Protein and RNA Chaperones }
\date{\today}

\author{Changbong Hyeon$^1$}
\email{hyeoncb@kias.re.kr}
\affiliation{Korea Institute for Advanced Study, Seoul 02455, Korea}
\author{D. Thirumalai$^2$}
\email{dave.thirumalai@gmail.com}
\affiliation{Department of Chemistry, University of Texas at Austin, Austin, Texas 78712, USA}

\begin{abstract}
Molecular chaperones are  machines that consume copious amount of ATP to drive misfolded proteins or RNA  to fold into functionally competent native states. Because the folding landscapes of biomolecules with complex native state topology are rugged consisting of multiple minima that are separated by large free energy barriers, folding occurs by the kinetic partitioning mechanism according to which only a small fraction of the molecules reach the folded state in biologically viable times.  The rescue of such proteins and RNA require chaperones. Although the protein and RNA chaperones are profoundly different in their structure and action, the  principles underlying their activity to produce the folded structures can be understood using a unified theoretical framework based on iterative annealing mechanism (IAM). Our theory shows that both these machines have evolved to the maximize the production of the steady state yield on biological times. Strikingly, theory predicts that only at a moderate level of RNA chaperone activity is the yield of the self-splicing pre-RNA is maximized in \textit{in vivo}. 
\end{abstract}
\maketitle

\noindent {\textbf{SIGNIFICANCE}} Proteins and RNA that do not fold spontaneously  with sufficient yield require molecular chaperones to reach the native structures. ATP-consuming molecular machines GroEL and CYT-19, respectively,  facilitate the production of proteins and group I intron in \textit{in vivo}.  Although the molecular mechanisms of their functions are dramatically different, the theory based on the Iterative Annealing Mechanism (IAM) provides a quantitative description of the outcomes of experiments. IAM postulates that chaperones utilize the free energy of ATP hydrolysis to rescue misfolded structures by providing multiple opportunities to fold. By operating under non-equilibrium conditions both GroEL and CYT-19 have evolved to maximize the yield of the folded state on biological times.   
\\

\section{INTRODUCTION}

The thermodynamic hypothesis, formulated by Anfinsen \cite{Anfinsen73Science,Anfinsen75APC}, asserts that the information needed to spontaneously reach the  unique three-dimensional (3D) folded states of proteins is fully encoded in the primary sequence~\cite{Anfinsen73Science}. 
This expectation is borne out in numerous \textit{in vitro} experiments on proteins with simple topologies and is the basis of structure prediction \cite{Jumper21Nature} and protein design~\cite{huang2016coming}.   
In addition to the uniqueness of the native fold, rapid folding kinetics could be important for the high fidelity of biological functions, as already implied by experiments showing the recovery of RNaseH activity upon urea denaturation and subsequent renaturation upon removal of urea~\cite{anfinsen1954JBC}. However, not all proteins fold spontaneously~\cite{Todd96PNAS,Kiefhaber95PNAS} with sufficient yield. Such proteins are either long and/or have complex topologies in the folded states. As a result, they are kinetically trapped in deep minima from which transition to the folded state takes place on time scales that are too long to be biologically viable. 


Trapping in metastable native-like states is lot more common in RNA than in proteins. Indeed, the folding kinetics of even RNA with simple architecture, such as hairpins, occurs in stages that is tracked to the propensity to populate multiple intermediates~\cite{Hyeon08JACS,kuznetsov2008NAR}. 
In general, a typical spectrum of RNA shows that there are several low lying excitations that are easily accessible from the folded state. The free energy barriers separating such states from the folded state in ribozymes are sufficiently large the transition times could be hours or longer~\cite{Pan1997JMB,Treiber99COSB}. 
The rescue of such recalcitrant proteins and RNA requires molecular chaperones, which likely evolved over a billion or more years ago.     


We synthesize experimental and theoretical developments to create a unified framework, based on the Iterative Annealing Mechanism (IAM), for understanding the action of protein and RNA chaperones. For concreteness, we focus on the \textit{Escherichia coli} GroEL/ES machinery (referred to as chaperonins) and the seemingly unrelated RNA enzyme CYT-19. 
Although both of these systems have been separately investigated~\cite{ThirumalaiBookDoniach,Woodson10RNABiol,Rajkowitsch07RNABiol,Herschlag95JBC,Russell2013RNAbiology}, there are only a handful of works~\cite{chakrabarti2017PNAS,Goloubinoff18NCB,Barducci15COSB,thirumalai2020PROTSCI,song2022moderate} that have provided a common description by invoking non-equilibrium mechanisms. Let us briefly introduce these two molecular machines.  (1) Although the identities of the proteins that are helped by GroEL, belonging to the heat shock family (HSP60), during the cell cycle, are not precisely known, no more than approximately ten percent  of the \textit{E. Coli.} proteome~\cite{lorimer1996FASEBJ} can recruit the chaperonin machinery to assist in the folding process. It is worth noting that assisted folding occurs without any violating the tenets of the Anfinsen hypothesis. In other words, the sequence encodes the folding mechanism.   
Importantly, GroEL is a promiscuous machine that assists the folding of proteins whose native states are structurally unrelated~\cite{Stan06PNAS}, implying it is
blind to the architecture of the folded proteins. 
(2)  That proteins can facilitate the folding of RNA was known over forty years thanks to the pioneering, but often overlooked, study~\cite{Karpel1980Biochem}, which showed that tRNA  could be reconstituted upon addition of proteins that bind to single stranded regions of RNA, which are presumably exposed in misfolded sates. 
More recently, it has been shown definitely that the ATP-consuming  DEAD-box protein CYT-19, found in \textit{Neurospora crassa} and the related Mss 116 in yeast, have chaperone activity~\cite{Mohr02Cell,turcq1992protein}. 
Importantly, the absence of CYT-19 in mutants of \textit{Neurospora crassa} resulted in the accumulation of group I intron. 
Like GroEL, the fungi-derived CYT-19 is also promiscuous in the sense that it exhibits chaperone activity towards RNA from other organisms~\cite{Mohr02Cell,Huang05PNAS}. 
Despite considerable differences between these two machines, which we explain below, the IAM provides a single framework that quantitatively explains all the available experiments and provides a platform to anticipate universal features of chaperone activity. \\  
 

\noindent {\bf Kinetic Partitioning Mechanism (KPM). } 
The need for chaperones is illustrated by considering spontaneous folding of RNA and proteins with complex topology. The folding landscapes of these biomolecules are rugged \cite{Thirumalai96ACR,Chen00PNAS,dill1997levinthal,Hyeon03PNAS} consisting of multiple minima separated by large free energy barriers that are difficult to cross in biologically relevant times.  There are two sources of frustration. Energetic frustration, arising from competing interactions between distinct arrangements of the amino acid residues~\cite{BryngelsonPNAS87,Wolynes95Science,OnuchicPNAS98,Todd96PNAS} (or nucleotides~\cite{Thirum05Biochem}). As a result, not all favorable interactions at specific residue or nucleotide location can be simultaneously satisfied~\cite{Thirum05Biochem,Todd96PNAS}. 
On the other hand, topological frustration~\cite{Guo95Biopolymers,Thirumalai96ACR}, is caused by the incompatibility between stable structures formed
on local length scales and the global native fold.  
Because of the rough free energy landscape, long-lived misfolded native-like metastable states could readily form during the folding reaction. 
This is indeed the case in the folding of Rubisco~\cite{todd94science,Todd96PNAS} and \textit{Tetrahymena} ribozyme~\cite{Pan1997JMB}. 

The KPM, which follows from the description of the folding landscape,  provides a unified theory ~\cite{Guo95Biopolymers,Todd96PNAS,Thirum05Biochem,Thirumalai96ACR,Thirumalai97TCA} to describe folding of both RNA and proteins under non-permissive conditions in which the yield of the folded state is paltry.  In this picture, only a small fraction, $\Phi$, of molecules reaches the native state. 
The remaining fraction is trapped in one of the many minima for arbitrarily long times. The non-convexity of folding landscape engenders multiple parallel folding pathways, giving rise to folding kinetics that are best fit to a multi-exponential function, 
\begin{align}
P_N(t)=1-\Phi e^{-t/\tau_f}-\sum_i\phi_{s,i}e^{-t/\tau_{s,i}}. 
\label{eqn:PN}
\end{align}
Here,  $\tau_f$ is the time for reaching the native state by the fast route, and $\phi_{s,i}$ and $\tau_{s,i}$ are those associated with slow routes, and the relations $\tau_{s,i}\gg \tau_f$. Conservation of flux implies that 
$\Phi+\sum_i\phi_{s,i}=1$ should be satisfied. 

\begin{figure*}[ht!]
\includegraphics[width=0.8\linewidth]{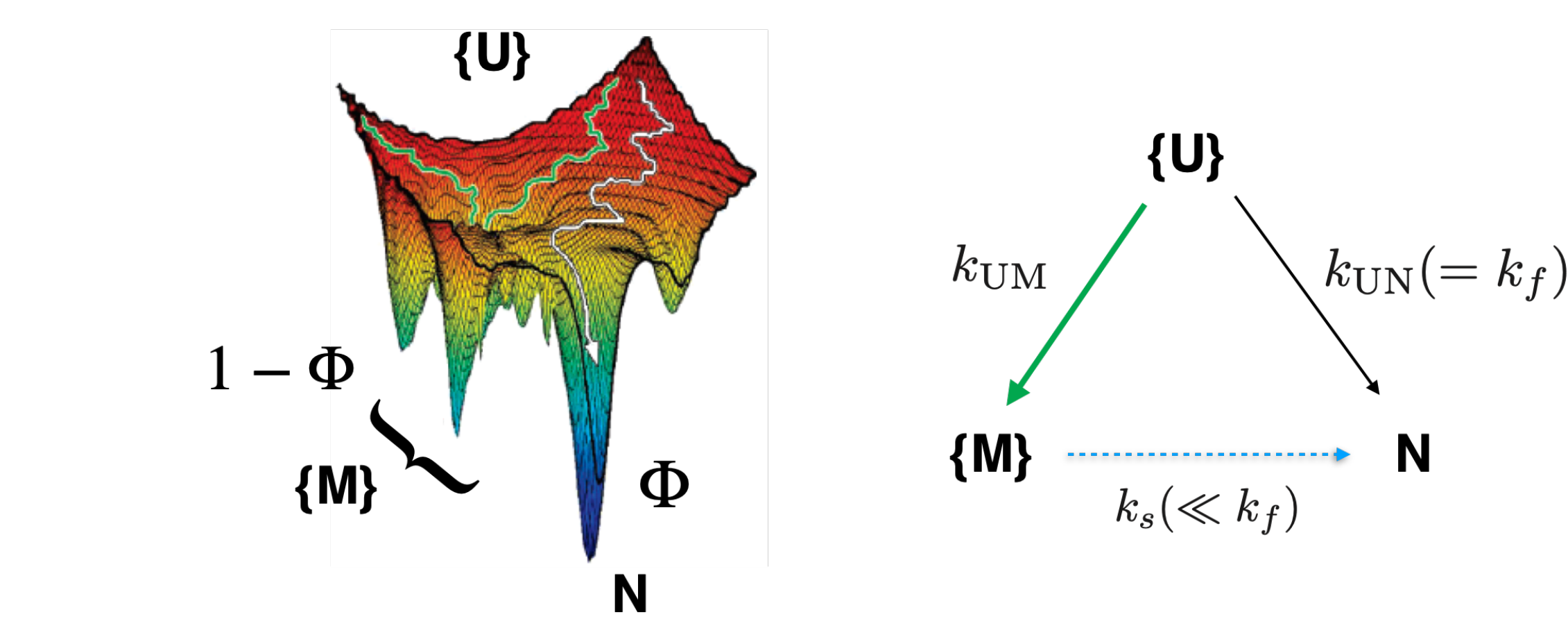}
\caption{Rugged folding landscape and the kinetic partitioning mechanism~\cite{Guo95Biopolymers}, highlighting the multiple
parallel pathways to the native state (N). The ensemble of unfolded states and folding intermediates
are denoted by {U}and {M}, respectively. The yield of the folded state during spontaneous is denoted by  the partition factor $\Phi$. The green lines are the slow trajectories. The fast trajectory in white with an arrow  reaches the folded state without being kinetically trapped in a deep local minimum. The three state reaction scheme on the right simplifies the KPM enabling an analytical solution.}
\label{fig:KPM} 
\end{figure*}

For small, fast-folding globular proteins, $\tau_f\sim \mathcal{O}(1)$ ms and $\Phi\approx 1$.  
However, this is not the case for proteins with complex topology, whose folding landscape is rugged as illustrated in Fig.~\ref{fig:KPM}. 
Navigation in such a landscape gives rise to predominantly slow track trajectories that result in KPM (Eq. \ref{eqn:PN}). 
With the  assumption that the ensemble of folding intermediates satisfies the condition, $\sum_i\phi_{s,i}=1-\Phi$ and $\tau_{s,i}\approx \tau_s$, $P_N(t)$ in Eq.~\eqref{eqn:PN} is simplifies as, 
\begin{align}
P_N(t)&\approx 1-\Phi e^{-t/\tau_f}-(1-\Phi)e^{-t/\tau_s}\nonumber\\
&\rightarrow \Phi(1-e^{-t/\tau_f}).  
\end{align}
The expression following the arrow follows if the folding time along the slow track, describing the transition from misfolded intermediates to the native state is prohibitively long, $\tau_{s}\gg \tau_f$. In this case, even if the stability of the native state is far greater than the intermediate state $\Delta G_{MN}=G_N-G_M\ll 0$, direct transitions from the misfolded intermediates to the native state are effectively forbidden on a biologically relevant time or on a time scale of \textit{in vitro}. 
As a result, $\Phi(1-e^{-t/\tau_f})\rightarrow \Phi$ for $t\gg \tau_f$, and only the fraction $\Phi$ of the entire population reach the native state. 

By assuming that the folding landscape in Fig.~\ref{fig:KPM} may be mapped onto a three-state model of $\{U\}$, $\{M\}$, and $N$, the \emph{partition factor}, $\Phi$, is determined by the ratio of initial collapse rates of an ensemble of unfolded conformations ($\{U\}$) into the native state ($N$) and misfolded intermediates ($\{M\}$). 
\begin{align}
\Phi=\frac{k_{\rm UN}}{k_{\rm UM}+k_{\rm UN}}. 
\label{eqn:partitioning}
\end{align}
Once trapped in long-lived kinetic intermediates in which the hydrophobic patches are not fully sequestered to the interior of the structure, 
functionally incompetent misfolded proteins accumulate, potentially forming protein aggregates, which are likely to be detrimental for the cell. Similarly, errors in base pairing or incorrect topological arrangement of reformed helices would result in metastable states with long lifetimes.

Several \textit{in vitro} experiments on RNA and proteins have been interpreted in terms of the KPM. (i) Folding of hen egg white  lysozyme near room temperature and neutral pH showed that $\Phi = 0.15$~\cite{Kiefhaber95PNAS}, can be increased to $\Phi =0.25 $~\cite{Matagne1998JMB} by changing pH. Thus, $\Phi$ can be increased or decreased not only by changing the external conditions but all also by mutations. (ii) The value of $\Phi$ for Rubisco is (0.02 - 0.05)~\cite{Tehver08JMB}, which makes it a stringent substrate for GroEL. (iii) The first ensemble experimental measurements on \emph{T.} ribozyme showed that $\Phi \approx 0.08$~\cite{Pan1997JMB},  which was subsequently  confirmed in single molecule experiments~\cite{Zhuang00Science}. Interestingly, upon stabilization of a point mutation that stabilized the P3 pseudoknot using a single point mutation increases by a factor of eight~\cite{Pan00JMB}.\\

\noindent {\bf GroEL-GroES machinery.}
To facilitate folding and maintain protein homeostasis, cells overexpress molecular chaperones, particularly under conditions that trigger aggregation, such as during heat shock.   
Although known in the context of genetics of bacteriophage assembly 
 and the synthesis of the large
subunit of Ribulose-1,5-bisphosphate carboxylase/oxygenase (Rubisco) over forty years ago~\cite{Friedman84MicrobiolRev,Milos84JCellBiochem}, it was only in the late 1980s~\cite{Lorimer89Nature,Ostermann89Nature} that it became clear that GroEL-GroES machinery is involved in the rescue of heterologously expressed proteins. 
GroEL, being a barrel-shaped heptameric oligomer with a seven fold symmetry with a large central cavity~\cite{Sigler98AnnRevBiochem}, naturally suggests that the substrate protein is simply sequestered in the Anfinsen cage till the native state is reached~\cite{Horwich09FEBSLett,Hoffmann10PNAS}. Recent developments have demonstrated that the neither the passive nor the active cage Anfinsen model, which argues that the protein folding rate is accelerated by confinement effects~\cite{BrinkerCell01,Gupta14JMB}, is relevant in the function of GroEL.  
Several experiments have altered our view of the function of GroEL, further supporting the theory based on IAM. (1) When challenged with  a misfolded substrate protein and supply of ATP, the GroEL machinery springs to action by undergoing large conformational changes, by twisting the barrel and doubling its  volume of cavity~\cite{Thirumalai01ARBB,Hyeon06PNAS}.  These allosteric transitions, requiring ATP binding and hydrolysis~\cite{Horovitz01JSB,HorovitzJMB94}, are required in helping proteins fold, implying the GroEL/ES plays an active role in its function. (2) Furthermore, ATP-binding and hydrolysis exert mechanical stress to the encapsulated proteins, which destabilizes the misfolded proteins~\cite{korobko2022diminished,koculi2021retardation,korobko2020Elife}. (2) Thanks to experiments from several laboratories it is firmly established that the symmetric (or ``American football" ) complex is the functional unit~\cite{todd94science,Sameshima10BiochemJ,Takei12JBC,Ye13PNAS,Yang13PNAS,Llorca94FEBSLett,Corrales96FD,Fei13PNAS,Fei14PNAS}.These experiments show that in the presence of folded proteins 
the GroEL/ES nanomachine becomes a parallel-processing machine, discharging all the ligands with each round of catalytic cycle, which maximzes the number of iterations (by minimizing the residence time of the substrate proteins in the cavity), thus enhancing the production of the native material, as predicted by the IAM (see below). 

\begin{figure*}[ht!]
\includegraphics[width=0.8\linewidth]{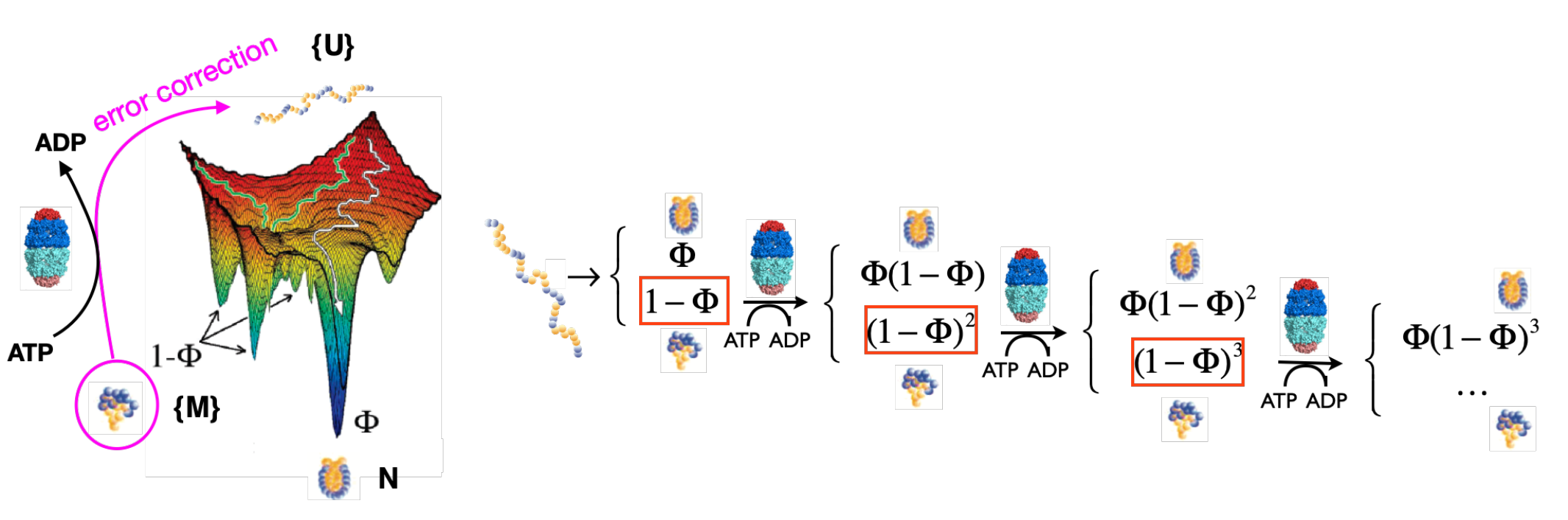}
\caption{Changes in the subpopulations of protein while the protein conformations are iteratively annealed to their native state through the GroEL-GroES chaperonin-assisted folding. As illustrated, in the symmetric functional unit the substrate proteins can be present in both the chambers, thus making GroEL/ES a parallel processing machine. The residence time in the cavity is minimized in the presence of the substrate protein.}
\label{fig:IAM} 
\end{figure*}

The action of GroEL conformation on the $(1-\Phi)$ population of misfolded proteins ($\{M\}$) is to  unfolds (at least partially) them to the $\{U\}$ state and offer another chance to refold. It is worth noting that when encapsulated in the cavity, the substrate protein undergoes partitioning rapidly, with certain probability, on time scales that is less than substrate protein residence time in the Anfinsen cage. In other words, when the substrate protein folds it does so in the cavity.
In the second round of folding, $\Phi(1-\Phi)$ would fold into the native state, and $(1-\Phi)^2$ would again misfold. 
When this process is repeated $n(= t/\tau_0)$ times, where $\tau_0$ is the time associated with a single cycle of chaperone action, which has been shown to be $\tau_0\approx 2$ sec, and $t$ is the time duration of the entire process, the fraction $(1 -\Phi)^n$ still remain misfolded, and hence, the $1 - (1 -\Phi)^n$ are folded. 
Thus, the yield of native state after $n$-cycle of chaperone action is obtained as 
\begin{align}
P_N(n)=1-(1-\Phi)^n\xrightarrow[n=t/\tau_0]{\Phi\ll 1}1-e^{-\Phi t/\tau_0}. 
\label{eqn:IAM_simple}
\end{align}
This is the gist of \emph{iterative annealing mechanism} of chaperone assisted protein folding, which is reminiscent of the simulated annealing protocol~\cite{ThirumalaiBookDoniach} for solving the optimization problem in computer science~\cite{Kirkpatrick83Science} and the stochastic resetting in target-search processes~\cite{evans2011diffusion,pal2023thermodynamic}. Then, if there are equal amount of GroEL/ES chaperonin system and proteins in the cell, the time required for a GroEL/ES particle to produce the 100 \% population of native states of the Rubisco  with $\Phi\approx 0.05$ is around 40 sec. 

Here, it is noteworthy that the steady state yield of native state is 100 \%, i.e., $P_N(n\rightarrow\infty)=1$. 
The resulting situation is fundamentally different from an expectation based on the landscape picture at thermal equilibrium, where the native state yield should be dictated by the Boltzmann distribution, $P_N^{eq}=1/(1+e^{-\Delta G_{\rm MN}/k_BT})$. 
One might argue that for sufficiently large stability of the native state, $\Delta G_{\rm MN}\ll 0$, 
such that folding landscapes are funneled, 
the native yield may be approximated to the unity ($P_N^{eq}\lesssim 1$). 
However, the steady state expression of the IAM is not identical to the Boltzmann distribution, $P_N(n\rightarrow\infty)\neq P_N^{eq}$. 
Given that  the operation of GroEL chaperonin cycle requires free energy consumption of $\sim(3-4)$ ATP molecules per cycle~\cite{Yang13PNAS}, 
the chaperone-assisted folding of proteins over the population 
should be viewed as being  far-from-equilibrium process where free energy is continuously injected and dissipated from the system~\cite{Mugnai2020RMP,Song2021JCP,Frank10PNAS}. 
The distinction of $P_N(\infty)$ from $P_N^{eq}$, which may not immediately be clear for GroEL-assisted folding of a protein with large stability, is further clarified by discussing  RNA chaperones 
whose action on RNA molecules is  different from GroEL,  RNA chaperones can destabilize the RNA conformation in the native state as well as the misfolded conformation. 
This aspect will be discussed in the next section. 
\\

\begin{figure*}[ht!]
\includegraphics[width=0.6\linewidth]{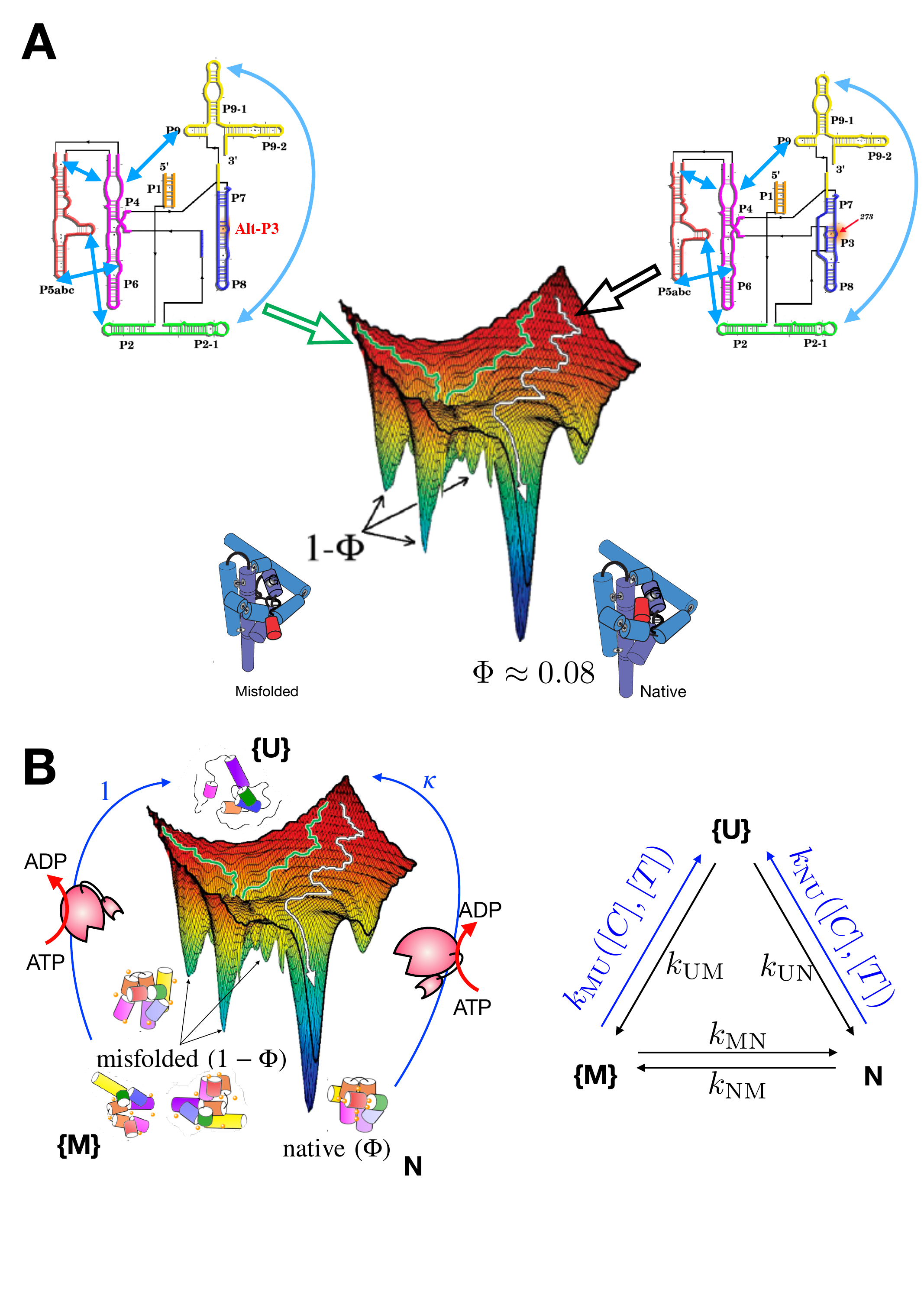}
\caption{. {{{\bf A}. Illustration of \emph{in vitro} folding of \emph{T. ribozyme} via kinetic partitioning. One of the major cause of misfolding due to the formation of Alt-P3 helix in the secondary structure is highlighted together with the differing pathways on the folding landscape. The partition factor is only $\Phi\approx 0.08$. 
{\bf B}. A folding landscape highlighting the role of ATP-burning RNA chaperone, CYT-19, and its mapping onto the 3-state kinetic model.}}Illustration of \emph{in vitro} folding of \emph{T. ribozyme} via kinetic partitioning. One of the major cause of misfolding due to the formation of Alt-P3 helix in the secondary structure is highlighted together with the differing pathways on the folding landscape. The partition factor is only $\Phi\approx 0.08$. 
{\bf B}. A folding landscape highlighting the role of ATP-burning RNA chaperone, CYT-19, and its mapping onto the 3-state kinetic model.}
\label{fig:RNA_chaperone} 
\end{figure*}

\noindent {\bf RNA chaperones.} 
Just like proteins, the folded state  of RNA,  is encoded in the primary sequence. Although there are exceptions, it is thought that RNA folding hierarchical that begins by first forming a secondary structure, followed by establishment of tertiary contacts between preformed secondary structural motifs~\cite{Thirum05Biochem}. The catalytic (self-cleavage) activity of
\emph{T}. ribozyme, which is one of the most extensively studied model systems for RNA folding,  requires that it be folded correctly. However, experiments complemented by theoretical arguments based on KPM show that only $\Phi <  0.1$ for molecules reach the folded state spontaneously, which means that the majority of molecules are kinetically trapped~\cite{Pan1997JMB}. For instance, the partition factor $\Phi$ is as small as 0.08 during \emph{in vitro} folding of \emph{T. ribozyme}~\cite{Pan1997JMB}. 
Incorrect formation of these structural motifs gives rise to a functionally incompetent ribozyme that compromises its cleavage activity. 
Fig.~\ref{fig:RNA_chaperone}A shows the two secondary structure maps of RNA. 
For this structure to properly self-assemble, five tertiary contacts between the preformed helices (indicated by aqua colored arrows) should be correctly form. 
In particular, the pseudo knot formation, defined by the P3 helix, is critical. 
Without the P3 helix, the  two major domain cannot be consolidated. Often during the folding process, alternative helix (Alt-P3) is formed resulting is a topological trap, which structure is  catalytically inactive~\cite{Pan00JMB}.

Because $\Phi$ is small, folding of RNAs such as \emph{T}. ribozyme requires chaperones \textit{in vivo}. DEAD-box protein CYT-19 is one of the general RNA chaperones, which belongs to superfamily-2 (SF2) helicase family, consisting of core helix domains and Arginine rich C-terminal tail~\cite{turcq1992protein,Mohr02Cell,Grohman2007Biochemistry,Russell2013RNAbiology,jarmoskaite2021atp}. If RNA have a surface-exposed helices or tertiary interaction motifs, as would be the case in the misfolded states, CYT-19 could recognize them and unwind the duplex into single strands. The unwinding process requires ATP consumption, which implies that CYT-19 activity is a non-equilibrium process. 

What distinguishes CYT-19 from GroEL is that CYT-19 can not only deactivate the native ribozyme but also unwind the surface exposed helices of misfolded riboyzme. 
Under less stabilizing condition, ribozymes show low cleavage activity as a result of deactivation by CYT-19, and this trend is amplified at higher CYT-19 concentration~\cite{Bhaskaran07Nature}. 
Such scenario of deactivation of native state is not included in the IAM of GroEL in Fig.~\ref{fig:IAM}. 
It was also found that the ribozyme at steady state ($t\rightarrow \infty$) in the presence of CYT-19 does not necessarily reach the unity as is predicted for GroEL assisted folding yield of protein. 

By considering an extent of chaperone-mediated disruption of native riboyzme relative to the misfolded ribozyme in terms of a factor $\kappa$ ($0\leq \kappa\leq 1$), 
one can follow a similar line of reasoning, so as to derive the RNA chaperone-assisted folding yield of native ribozyme after the $n$-th round of chaperonin action: 
\begin{align}
P^{\kappa}_N(n)=\Phi\frac{1-(1-\kappa)^n(1-\Phi)^n}{\kappa+(1-\kappa)\Phi}. 
\label{eqn:IAM_general}
\end{align}
This is the generalized expression of native yield by chaperone-assisted folding of biopolymers. 
The steady-state yield of the native state is give by 
\begin{align}
P^{\kappa}_N(\infty)=\frac{\Phi}{\kappa+(1-\kappa)\Phi}.
\label{eqn:IAM_ss}
\end{align}
Note that Eq.~\eqref{eqn:IAM_simple} is recovered by setting the relative disruption factor $\kappa=0$ in Eq.~\eqref{eqn:IAM_general}. 
In fact, the landscape model depicted in Fig.~\ref{fig:RNA_chaperone}B can be mapped onto the 3-state kinetic model involving $\{U\}$, $\{M\}$ and $N$, by incorporating the chaperone-mediated reverse transitions from $\{M\}$ to $\{U\}$ and from $N$ to $\{U\}$, which are missing in Fig.~\ref{fig:KPM}.   
Such a model allows one to express the steady-state ($t\rightarrow\infty$ or $n\rightarrow \infty$) yield of native state in terms of 
the six rate constants. 
Further, under the condition relevant to our discussion of chaperone-assisted folding of biopolymers, i.e., $k_{\rm NM}$, $k_{\rm MN}\ll 1$ and $k_{\rm UN}\gg k_{\rm NU}$, it is straightforward to obtain 
\begin{align}
P_N^\kappa(\infty)\approx \frac{k_{\rm UN}}{\left(\frac{k_{\rm NU}([C],[T])}{k_{\rm MU}([C],[T])}\right)k_{\rm UM}+k_{\rm UN}}. 
\label{eqn:PN_inf}  
\end{align} 
Inserting the expression of the partition factor $\Phi$ (Eq.~\eqref{eqn:partitioning}) into Eq.~\eqref{eqn:IAM_ss} 
gives $P_N^\kappa(\infty)\approx \frac{k_{\rm UN}}{\kappa k_{\rm UM}+k_{\rm UN}}$. 
Thus, the factor $\kappa$ turns out to be the ratio of chaperone-induced unfolding rate from native and misfolded state. 
\begin{align}
\kappa=\frac{k_{\rm NU}([C],[T])}{k_{\rm MU}([C],[T])}, 
\label{eqn:kappa}
\end{align} 
where $[C]$ and $[T]$ in the argument denote the chaperone and ATP concentrations, respectively, thus making explicit the dependence of the chaperone-induced unfolding rates on them. 

Although the expression is simple, 
Eq.~\eqref{eqn:PN_inf} along with Eq.~\eqref{eqn:kappa} is of great significance. 
First, steady-state yield of native state is less than 1 unless $\kappa$ is zero, which is consistent with the Bhaskaran \emph{et al.}'s measurement~\cite{Bhaskaran07Nature}. 
Second, the relative disruption factor $\kappa$ is the ratio of two unfolding transition rates, both of which depend on the concentrations of chaperone and ATP. 
In the absence of ATP or chaperone ($k_{\rm NU}$, $k_{\rm MU}=0$), 
the majority of \emph{T.} ribozymes ($1-\Phi\approx 0.92$) are trapped in the misfolded conformations devoid of catalytic power.  
Given the fact that $\sim 100$ ATP molecules are consumed to fully refold a single misfolded ribozyme~\cite{jarmoskaite2021atp,song2022moderate}, one cannot stress enough the outcomes arising from the energy-expending, nonequilibrium nature of chaperone-assisted folding. \\





\begin{figure*}[ht!]
\includegraphics[width=1.0\linewidth]{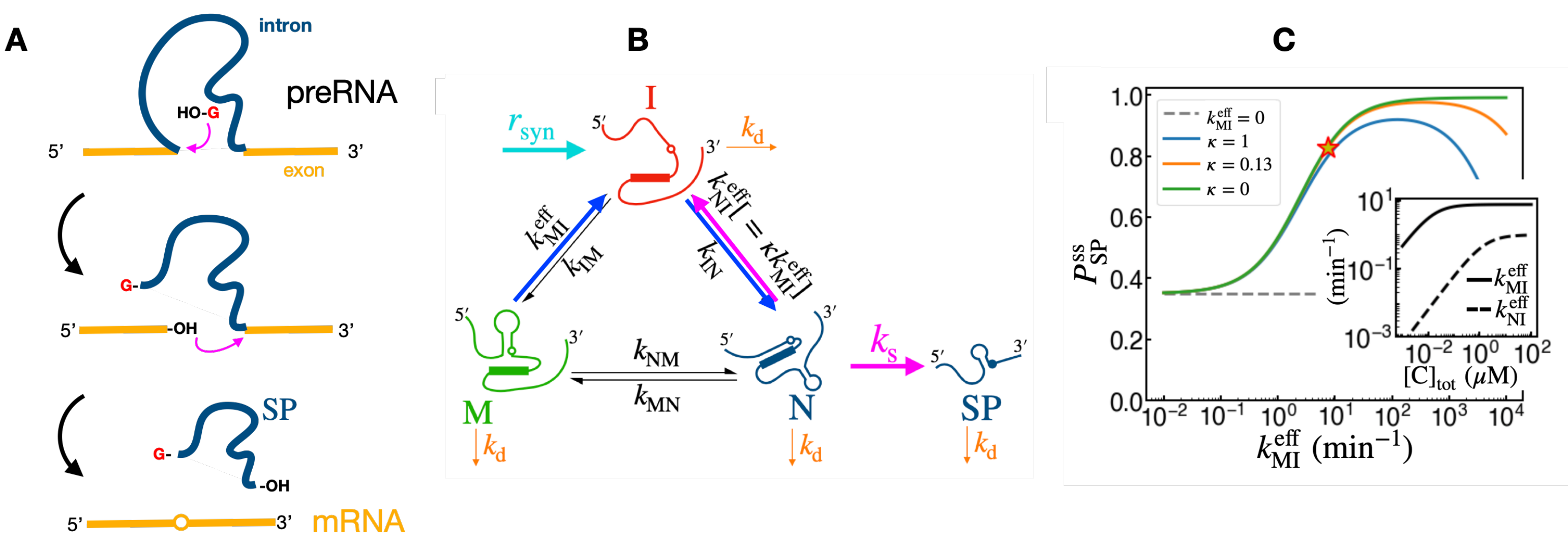}
\caption{{{\bf A}. Schematic  of group I intron splicing of preRNA leading to the production of mRNA transcript. {\bf B}. Kinetic network model combining chaperone-mediated folding
and the self-splicing of pre-RNA. RNA degradation rate is $k_d$ that could occur from all the relevant states. Experiments \cite{Mohr02Cell,Koduvayur04RNA,Jackson06RNA} suggest that $k_s$ is greater than $k_{I \rightarrow M}$ and $k_{I \rightarrow N}$  and $k_d < k_{I \rightarrow M}, k_{I \rightarrow N} < k_s$ where $k_s$ is the splicing rate.
{\bf C}. The yield of the self-spliced preRNA as a function
of chaperone activity ($k_{\rm eff}^{\rm 
MI}$), at different values of the recognition factor, $\kappa$. The rate constants ($k_{I \rightarrow M}$, $k_{M \rightarrow N}$, $k_{N \rightarrow M}$, $k_d$, and $k_s$) used in the kinetic model ({\bf B}) are listed in the caption to Fig. 5{\bf B} in \cite{song2022moderate}. The star symbol is evaluated at $k_{\rm MI}^{\rm eff} \approx 7.5$ min$^{-1}$ (the saturation value in inset) and $\kappa = 0.13$.  }}
\label{fig:invivo} 
\end{figure*}

\noindent {\bf RNA folding in cells. }
Self-splicing reaction involving the removal of the the non-coding intron occurs in the absence of proteins. However,  RNA must be folded, which as described above, requires DEAD-box proteins. To account for RNA folding and the splicing reaction simultaneously, we created a theory~\cite{song2022moderate} by extending the IAM to assess  the effect of RNA chaperone on the yield of the self-spliced pre-RNA ($SP$), which is the main product of the catalytic activity of the group I intron \emph{T}. ribozyme (Fig.~\ref{fig:invivo}A).  
Figure~\ref{fig:invivo}B shows the kinetic network that describes the interplay between the folding dynamics of \emph{T}. ribozyme, chaperone-mediated unfolding, and self-splicing reactions as well as the degradation of the transcript. 
Pre-RNA, synthesized during transcription, first reaches the $I$ state, followed by  folding to the $N$ state or misfolding to the $M$ state of the ribozyme, as described by the KPM.  Self-splicing of the catalytically competent state of ribozyme ensues to yield the state denoted by $SP$.  
The self-splicing rate ($k_s$) is typically greater than the folding rates~\cite{Koduvayur04RNA,Jackson06RNA}. Moreover, folding occurs much faster than the degradation rates ($k_d\ll k_{\rm IM}$, $k_{\rm IN}\ll k_s$), which is assumed to be independent of the RNA conformation. 
Let the chaperone-mediated unfolding be $k_{\rm MI}^{\rm eff}$ and $k_{\rm NI}^{\rm eff}$. Given that  the transitions between $M$ and $N$ states are prohibitively slow $k_{\rm NM}\ll 1$ and $k_{\rm MN}\ll 1$, the expression of the steady state population of SP ($P_{\rm SP}^{\rm ss}$) is straightforward to derive, as described in detail in the supplementary information given in our theory~\cite{song2022moderate}. 

Figure~\ref{fig:invivo}C shows the chaperone activity-dependent steady-state yield of $P_{\rm SP}^{\rm ss}$ for various values of the recognition factor, $\kappa$. 
Of particular note is that for $\kappa\neq 0$, 
$P_{\rm SP}^{\rm ss}$ varies non-monotonically with the unfolding activity of RNA chaperones ($k_{\rm MI}^{\rm eff}$), reaching a  maximum at,
\begin{align}
k_{\rm MI}^{\rm eff}\approx \sqrt{\frac{k_{\rm IM}k_s}{\kappa}},
\label{effective}
\end{align} 
which is equivalent to $k_{\rm MI}^{\rm eff}k_{\rm NI}^{\rm eff}\approx k_{\rm IM}k_s$. 
This shows that RNA splicing is maximized at only a moderate level of RNA chaperone activity. 
To attain  the maximum yield of $SP$, 
the chaperone activity should be large enough to rescue pre-RNA from the misfolded states ($k_{\rm MI}^{\rm eff}>k_{\rm IM}$), 
while the chaperone-mediated unfolding of native state must not overwhelm the rate of splicing 
($k_{\rm NI}^{\rm eff}<k_s$). \\

\section{DISCUSSION}

\noindent {\bf Non-equilibrium  effects: }
We showed previously that the steady values of the yield of the native states of CYT-19 mediated  folding of \emph{T}. ribozyme and its variant as well as GroEL-assisted reconstitution of Rubisco and MDH do not approach the Boltzmann distribution (see Figs (4-6) in~\cite{chakrabarti2017PNAS}).  This implies that chaperones drive the substrate population out of equilibrium~\cite{chakrabarti2017PNAS}, thus, increasing accessibility of the native state on a biologically relevant time scale. 
What matters for the cellular function is the supply rate of native enzyme production ($P_N/\tau$) that can match the cellular demand, not the yield of native state ($P_N$). 
The situation is analogous to the tradeoff between  efficiency of heat engines (Carnot engine for example) and generation of maximal power. 
Although maximum Carnot efficiency is achieved when operated under quasi-static conditions, power generation would be compromised. 
From the perspective of power generation, it is more important to operate the engine at a fast rate (non-equilibrium conditions) than in the quasi-static limit 
where the thermodynamic efficiency is maximized,  with zero power.  This well-known concept (power-efficiency tradeoff), in  macroscopic heat engines, is also valid in microscopic  engines in stochastic environment~\cite{shiraishi2016PRL,pietzonka2018universal,dechant2018PRE,Mugnai2020RMP}, and is also found in the function of chaperones, which existed several billions years before steam engines were invented.
It is clear that as long as cells tolerate misfolded enzymes (errors), execution of biological functions is determined by the production rate of catalytically active enzymes even at the expense of lavish consumption of ATP. 
\\

\begin{figure*}[ht!]
\includegraphics[width=0.6\linewidth]{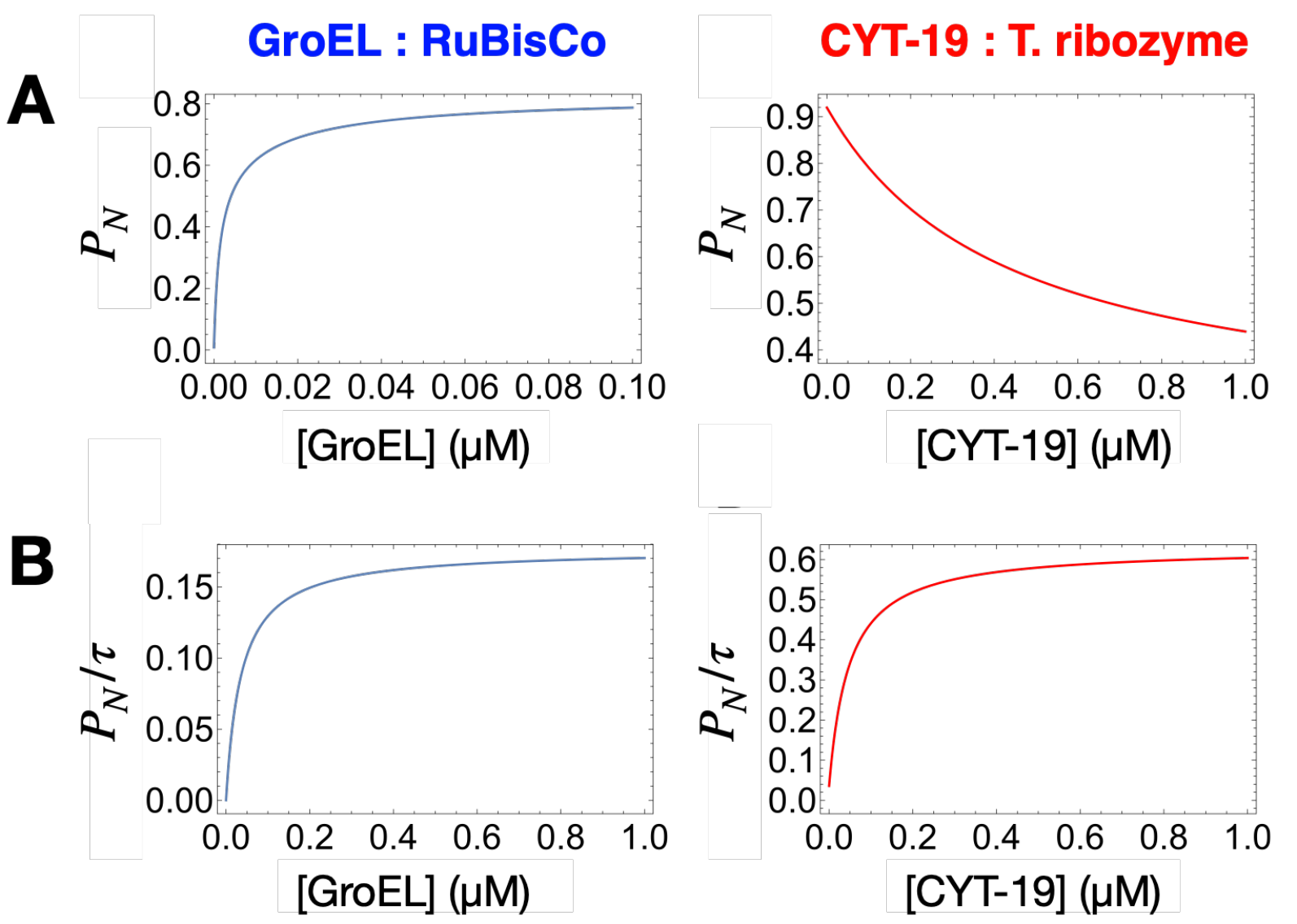}
\caption{{\bf A}. Steady-state yield ($P_N$) of RuBisCo (left) and T. ribozyme (right) as a function of the corresponding chaperone concentration. 
{\bf B}. The steady-state yield per unit relaxation time of chaperone-assisted folding ($P_N/\tau$) 
as a function of the chaperone concentration. In both cases, $P_N/\tau$ is an increasing function of the chaperone concentration in contrast to $P_N$ in {\bf A}.}
\label{fig:yield_dynamics} 
\end{figure*}

\noindent {\bf Chaperones solve an optimization problem: } A general question that arises in the functions of biological motors is what is being optimized given the available resources in the cell~\cite{Mugnai2020RMP}.  It is difficult to answer  this question in complete generality because various biological machines have evolved to perform myriad of totally unrelated functions. For instance, in the case of the well studied molecular motors that ferry cargo (kinesin, myosin, and dynein) by walking on polar tracks one could wonder if evolution has optimized processivity or speed. Although concepts from non-equilibrium statistical physics, especially thermodynamic uncertainty relations~\cite{Hwang2018JPCL,Song2021JCP,barato2015PRL,seifert2019stochastic} and information theoretic approaches~\cite{Leighton24ARPC} have given insights, the answer might depend on the system under consideration.  
However, in the context chaperones, one can make a clear prediction that is supported by experimental data.  
This is vividly illustrated in Figure~\ref{fig:yield_dynamics}, which contrasts the steady state yields of Rubisco and \emph{T}. ribozyme when the concentration of  GroEL and CYT-19 are varied.
In the presence of GroEL, which preferentially recognizes the misfolded states of proteins,  the steady state yield ($P_N$) of native Rubisco increases as the concentration of GroEL is increased. In sharp contrast, increase in the concentration of CYT-19 reduces $P_N$ of the functional \emph{T.} ribozyme.  
The solution to the master equation~\cite{chakrabarti2017PNAS,song2022moderate} associated with the reversible 3-state model (Fig.~\ref{fig:RNA_chaperone}B) indicates that these opposing trends originate from the distinct actions of the two chaperones on Rubisco and 
\emph{T.} ribozyme, which is quantified with different value of $\kappa$ (Eq.~\ref{eqn:kappa}).  
The value of  $\kappa\approx 0$ for GroEL on Rubisco and $\kappa\approx 0.13$ for CYT-19 for \emph{T.} ribozyme~\cite{song2022moderate}. 
Strikingly, the IAM theory predicts that $P_N/\tau$ ($\tau$ is the longest relaxation time ($\approx$ the folding time) needed to reach the steady state) is an increasing function of the chaperone concentration for both GroEL and CYT-19 (Fig.~\ref{fig:RNA_chaperone}B). In other words, for a given concentration of chaperone, the product of the steady-state yield and the folding rate serves as a good indicator of the action of chaperone.    
 \\

\noindent \textbf{Lavish consumption of ATP. } It is important to calculate the thermodynamic cost incurred in the process of annealing the structures. Both the machines consume copious amount of ATP in the process of rescuing misfolded structures. The theory shows about 125 ATP molecules are consumed to drive a single \emph{T.} ribozyme depending on the concentration of CYT-19~\cite{song2022moderate}. Similarly, GroEL/ES systems burns between  (60-120) ATP molecules in order to fold a single Rubisco molecule~\cite{Todd96PNAS}. Not withstanding argument that this is fraction of the cost needed to synthesize the protein and RNA, it has to be concluded that these machines are inefficient. 
These findings suggest that to maintain cellular homeostasis, minimization of error in protein and RNA functionality is prioritized over the high thermodynamic cost.\\

\noindent\textbf{Evolutionary Considerations:} It is interesting to assess the potential evolutionary implications in light of several experiments~\cite{Fayet89JBacteriology,Friedman84MicrobiolRev,Fayer86MolecularMGG}. Of interest here are the early experiments, which showed that overexpression of GroEL facilitated assembly of heterologously expressed Rubisco
subunits~\cite{Van89Nature}. Similarly, overexpression of GroEL/ES suppressed effects of temperature sensitive  mutants of coat protein~\cite{Gordon94JBC}, by enhancing folding efficiency and the stability of the protein, which could have been an early demonstration of non-equilibrium effects~\cite{chakrabarti2017PNAS}. More recently~\cite{Bershtein13MolCell}, 
using experiments combined with a kinetic model showed that chaperonins rescue the growth of \textit{E. Coli.} deleterious DHFR mutants.  
In particular, the growth rate of the mutant strains was 
correlated with overabundance of folded DHFR, which in turn depended on GroEL/ES concentration.  In formulating the IAM~\cite{Todd96PNAS}, we proposed that the GroEL/ES machinery could positively affect potentially deleterious effects of mutations. 
In case of deleterious mutations, the partition factor would decrease. The ability of chaperonins to rescue these conformers and allow them multiple
chances to advance to the folded state explains experimental observations~\cite{Van89Nature,Gordon94JBC}. 
Evolution also searches for new biological activities by mutating 
existing genes. 
In such cases, chaperonins would ensure that a mutant protein would explore the folding landscape by avoiding trapping in deep local minima. 
When sufficient mutations accumulate  an earlier kinetic trap could evolves to a  new energy minimum with an altered function. 
These considerations suggest that chaperonins, and the optimization of folding by IAM 
must have been a significant early evolutionary event.
\\

\section{CONCLUDING REMARKS} 
We have presented a common theoretical framework to quantitatively account for the functions (facilitate the folding of proteins and RNA that cannot do so spontaneously) of two seemingly unrelated molecular machines that must have appeared early in evolution. There are a couple of differences between GroEL/ES and CYT-19, the two case studies used to illustrate the theory. (1) The annealing action of GroEL arises as a result of the changes in the microenvironment that the substrate protein experiences as result of ATP and GroES binding. Not only do these events double the volume of the central cavity but the cavity changes from being hydrophobic before binding to hydrophilic after binding. It is the change in the microenvironment that serves as the mechanism of annealing action of GroEL. It is likely that out theory  is allicable to the function of mitochondrial Hsp60 whose architecture is strikingly similar to GroEL/ES machine~\cite{Braxton24NSMB}. (2) CYT-19 interacts with single stranded or unstable regions of RNA, which occur readily when it misfolds. The helicase activity unfolds the helices in an ATP dependent manner, thus altering the folding landscape of the RNA. In this sense, CYT-19 is more actively engaged in interaction with misfolded structures than GroEL.  During each catalytic cycle, the unfolding process places the RNA in different regions of the rugged folding landscape from which folding can commence anew. Unlike GroEL, the structural basis of ATP binding, hydrolysis rates, the mechanism of release of the RNA (folded or not), and the extent of processivity of the helicase are not known. (3) The most striking difference between the two motors lies in the interaction with the misfolded structures. GroEL predominantly recognizes only misfolded structures in which the hydrophobic residues are exposed. In sharp contrast, CYT-19 interacts with both misfolded as well as the native state. The less stable misfolded structures are recognized by CYT-19 with greater probability than the folded state.    

Despite these differences, the theory based on the IAM quantitatively accounts for all the experimental observations~\cite{chakrabarti2017PNAS}. A striking prediction of the theory is that the action of protein and RNA chaperones may be universal in that they both maximize the native state yield on biological times by operating out of equilibrium.  Importantly, extension of the theory predicts that, under cellular conditions,
chaperone-mediated folding of group I intron ribozyme and the self-splicing reaction compete. Thus, theory predicts that to maximize the yield of the cleavage reaction, the chaperone should disrupt the conformations of the unfolded structures but not the native state before the splicing reactions. This dual requirement places bounds on the differences between the stability native and misfolded structures and the various kinetic rates in Figure \ref{fig:invivo}. Our prediction awaits experimental tests. Finally, we note that DEAD-box protein CsdA was shown to accelerate ribosome assembly by means of IAM~\cite{sun2025disassembly}, thus establishing the generality of the theory.
\\

\noindent \textbf{Acknowledgment:}
One of us (DT) is grateful to Erich Sackmann for sharing an office during his visits to Technical University of Munich. Although ES had retired, he remained a teacher, exposing DT to several interesting problems in Biophysics. We are indebted to George H. Lorimer, Shaon Chakrabarti, and Xiang Ye for fruitful collaboration. CH thanks the Center for Advanced Computation in KIAS for providing the computing resources. DT  was supported by a grant from the National Science Foundation (CHE 2320256),  the Welch Foundation through the Collie-Welch Chair (F-0019), and the US-Israel Binational Grant (2021077). 

\noindent \textbf{Declaration of Interests form:} The authors declare no competing interests.



\bibliographystyle{unsrt}

\end{document}